\documentclass[12pt,aps,prd]{revtex4}
\usepackage{amsmath}
\usepackage{graphicx}

\newcommand{\eq}[1]{Eq.~\eqref{eq:#1}}

\begin{document}
\title{Lattice Gauge Fixing as Quenching 
and the Violation of Spectral Positivity}

\author{C. Aubin}
\author{Michael C. Ogilvie}

\affiliation{Washington University, St. Louis, MO 63130}

\begin{abstract}
Lattice Landau gauge and other related lattice gauge fixing schemes
are known to violate spectral positivity.  The most direct sign of the
violation is the rise of the effective mass as a function of distance.
The origin of this phenomenon lies in the quenched character of the
auxiliary field $g$ used to implement lattice gauge fixing, and is
similar to quenched QCD in this respect.  This is best studied using
the PJLZ formalism, leading to a class of covariant gauges similar to
the one-parameter class of covariant gauges commonly used in continuum
gauge theories.  Soluble models are used to illustrate the origin of
the violation of spectral positivity. The phase diagram of the lattice
theory, as a function of the gauge coupling $\beta$ and the
gauge-fixing parameter $\alpha$, is similar to that of the unquenched
theory, a Higgs model of a type first studied by Fradkin and
Shenker. The gluon propagator is interpreted as yielding bound states
in the confined phase, and a mixture of fundamental particles in the
Higgs phase, but lattice simulation shows the two phases are
connected. Gauge field propagators from the simulation of an $SU(2)$
lattice gauge theory on a $20^4$ lattice are well described by a
quenched mass-mixing model.  The mass of the lightest state, which we
interpret as the gluon mass, appears to be independent of $\alpha$ for
sufficiently large $\alpha$.
\end{abstract}

\maketitle

\section{Introduction}\label{sec:intro}

Although many interesting properties of a lattice gauge theory
can be determined without gauge fixing,
there are several reasons
why it is needed.
At a fundamental level,
gauge fixing is
necessary to make the connection between continuum and lattice gauge fields.
Gauge fixing has also been a key technique in
lattice studies of confinement as well \cite{Greensite:2003bk}, and may help
differentiate between different models of confinement. 
For example, continuum theories of the origin of confinement often make predictions about
the gauge field propagator; see Ref.~\cite{Mandula:nj} for a review. 
Finally, gauge fixing may be
needed to determine important properties of the quark-gluon plasma phase of
QCD, such as screening masses, which are contained in the finite-temperature
gluon propagator 
\cite{Heller:1995qc,Heller:1997nq,Cucchieri:2000cy,Nakamura:2003pu}.

Techniques for lattice gauge fixing have been known for some time 
\cite{Mandula:rh}.
It has been clear from the beginning that non-Abelian lattice
gauge field propagators show a violation of spectral positivity. In a theory
respecting spectral positivity, the K\"all\'en-Lehmann representation
for a typical wall-to-wall two-point function has the form
\begin{equation}
G(t)=\int_{0}^{\infty }dm^{2}\,\,\rho \left( m^{2}\right) \frac{1}{2m}
e^{-m\left| t\right| } 
\end{equation}
where $\rho$ is everywhere non-negative.  Wall sources set all momenta
transverse to the direction of propagation to zero.  The effective
mass defined by
\begin{equation}
  m_{eff}\left( t\right) =-\frac{d}{dt}\ln G(t) 
\end{equation}
can be thought of as an average mass $\left\langle m(t)\right\rangle$,
where the average is taken over $\rho\left( m^{2}\right)
\exp({-m\left| t\right| })/{2m}.$ It follows that
\begin{equation}
  \frac{d}{dt}m_{eff}\left( t\right)=\left\langle m(t)
  \right\rangle ^{2}-\left\langle m^{2}(t)\right\rangle \leq 0.
\end{equation}
The positivity of $\rho$ implies that the effective mass is a non-increasing
function of $t$ which goes to the lightest mass propagated in the limit 
$t\rightarrow \infty $.

Gauge invariant operators which couple to glueball states show this
behavior, but the gluon propagator does not. Covariant gauge gluon
propagators have an effective mass increasing with distance. This is
not too surprising: we know from perturbation theory that covariant
gauges contain states of negative norm. However, that knowledge has
neither explained the form of the lattice gluon propagator nor aided
in the interpretation of the mass parameters measured from it. In
fact, the particular form of spectral positivity violation observed
for non-Abelian models is not observed in the $U(1)$ case
\cite{Coddington:yz}, which also has negative-norm states in covariant
gauges.

In lattice simulations, gauge fixing has typically involved choosing a
particular configuration on each gauge orbit. A brief review of this
approach is given in Ref.~\cite{Mandula:nj}. In the continuum, on the
other hand, gauge fixing most often includes a parameter that causes
the functional integral to peak around a particular configuration on
the gauge orbit. The PJLZ formalism, a comparable formalism for
lattice gauge fixing, first appeared in
Refs.~\cite{Parrinello:1990pm,Zwanziger:tn}.  The strong-coupling
expansion was developed in Ref.~\cite{Fachin:1991pu} and the gluon
propagator was studied in Ref.~\cite{Henty:1996kv}.  This formalism was
used in a discussion of Abelian projection in lattice theories
\cite{Ogilvie:1998wu}. There it was shown that Abelian projection
without gauge fixing leads to an equality of the asymptotic string
tensions in the underlying non-Abelian theory and the projected
Abelian theory. Furthermore, the string tensions were proven to be
equal with gauge fixing, provided the gauge fixing procedure respects
spectral positivity. The failure of spectral positivity demonstrated
in Ref.~\cite{Aubin:2003ih} explains the string tension discrepancies noted
in lattice studies of Abelian projection.  A lattice study of the
phase structure of the projected theory as a function of the
gauge-fixing parameter was carried out in Ref.~\cite{Mitrjushkin:2001hr}.

Our aim in this work is to demonstrate that spectral positivity is
violated in lattice gauge fixing by mechanisms related to those
encountered in quenched QCD.  Specifically, we argue that the gluon
propagator has a double pole structure similar to that of the
$\eta'$. We show in Sec.~\ref{sec:lgfix_qu} that the PJLZ formalism
makes clear that lattice gauge fixing is a form of quenching, with the
gauge transformations acting as quenched fields.  In
Sec.~\ref{sec:qu_form}, we develop a general formalism for quenched
fields, and apply that formalism to gauge fixing. We show that in the
case of a $U(1)$ gauge theory, spectral positivity is
maintained. Sec.~\ref{sec:mix_models} examines two useful simple
models for quenched fields based on mass mixing between a quenched and
an unquenched field. The first model involves scalars; the second
involves vector fields, and will be used to fit the results of our
lattice simulations. In Sec.~\ref{sec:phase_diag}, we discuss the
phase diagram of the gauge-fixed model, using the interpretation of
gauge fixing as a quenched Higgs model. As first discussed by Fradkin
and Shenker, the nominal Higgs and confining phases of the unquenched
model are connected, and this carries over to the quenched version. We
propose interpretations for the lattice gauge field propagator in both
phases. Section \ref{sec:lat_results} discusses the simulation results
for $SU(2)$ lattice propagators. A comparison of our results with some
of the other proposed forms for the lattice gauge field propagator is
performed in Sec.~\ref{sec:disc}, and Sec.~\ref{sec:conc} gives our
conclusions.

\section{Lattice Gauge Fixing as Quenching}\label{sec:lgfix_qu}

The standard approach to lattice gauge fixing is a two-step process
\cite{Mandula:nj}.
The
gauge fields $U_{\mu }\left( x\right) $ are associated with links of the
lattice, and take on values in a compact Lie group $G$. An ensemble of
lattice gauge field configurations is generated using standard Monte Carlo
methods. This ensemble of G-field configurations is generated by a
functional integral 
\begin{equation}
Z_{U}=\int \left[ dU\right] e^{S_{U}\left[ U\right] }~. 
\end{equation}
$S_{U}$ is a
gauge-invariant action for the gauge fields, \textit{e.g.}, the Wilson
action for $SU(N)$ gauge fields: 
\begin{equation}\label{eq:wilson_action}
S_{U}=\frac{\beta }{2N}\sum_{plaq}\,Tr\,\left( U_{plaq}+U_{plaq}^{+}\right) 
\end{equation}
where $U_{plaq}$ is a plaquette variable composed from link variables, and
the sum is over all plaquettes of the lattice. The gauge action $S_{U}$ is
invariant under gauge transformations of the form 
\begin{equation}
  U_{\mu }\left( x\right) \rightarrow g^{+}\left( x+\mu \right) 
  U_{\mu }\left( x\right) g\left( x \right) . 
\end{equation}
The expectation value of any observable $O$ is given formally by 
\begin{equation}
\left\langle O\right\rangle =\frac{1}{Z_{U}}\int \left[ dU\right]
\,e^{S_{U}\left[ U\right] }\,O\,, 
\end{equation}
but in simulations is evaluated by an average over an ensemble of field
configurations: 
\begin{equation}
  \left\langle O\right\rangle =\frac{1}{n}\sum_{i=1}^{n}\,O_{i}\,. 
\end{equation}

In order to measure gauge-variant observables, each field configuration in
the $U$-ensemble is placed in a particular gauge, \textit{i.e.}, a gauge
tranformation is applied to each configuration in the $U$-ensemble which
moves the configuration along the gauge orbit to a gauge-equivalent
configuration satisfying a lattice gauge fixing condition. The simplest
gauge choice is defined by maximizing 
\begin{equation}
  \sum_{x,\mu }Tr\,\left[ U_{\mu }\left( x\right) +U_{\mu }^{+}\left( x\right)
\right] 
\end{equation}
for each configuration over the class of all gauge transformations. The sum
is over all the links of the lattice. Any local extremum of this functional
satisfies a lattice form of the Landau gauge condition: 
\begin{equation}\label{eq:landau_gauge}
  \sum_{\mu }\left[ A_{\mu }\left( x+\mu \right) -A_{\mu }\left( x\right)
    \right] =0 
\end{equation}
where $A_{\mu }\left( x\right) $ is a lattice approximation to the continuum
gauge field, given by 
\begin{equation}
  A_{\mu }\left( x\right) =\frac{U_{\mu }\left( x\right) -U_{\mu }^{+}\left(
    x\right) }{2i}-\frac{1}{N}Tr\left[ \frac{U_{\mu }\left( x\right) -U_{\mu
      }^{+}\left( x\right) }{2i}\right] . 
\end{equation}
A lattice form of Coulomb gauge can be obtained by restricting the sum
over $\mu$ in \eq{landau_gauge} above to the spatial dimensions. Other
gauge-fixing conditions may also be used \cite{Giusti:2001xf}, and
improvement can be applied to the definition of $A_{\mu }$ as
well. The global maximization needed is often implemented as a local
iterative maximization. The issue of Gribov copies arises in lattice
gauge fixing because such a local algorithm tends to find local maxima
of the gauge-fixing functional. There are variations on the basic
algorithm that ensure a unique choice from among local maxima
\cite{Giusti:2001xf}.

For analytical purposes, it is necessary to generalize this procedure 
\cite{Fachin:1991pu}, so
that a given single configuration of gauge fields will be associated with an
ensemble of configurations of $g$-fields. We will generate this ensemble
using 
\begin{equation}
  S_{gf}\left[ U,g\right] =\sum_{l}\frac{\alpha }{2N}Tr\,\left[ g^{+}\left(
    x+\mu \right) U_{\mu }\left( x\right) g\left( x \right) +g^{+}\left( x
    \right) U_{\mu }^{+}\left( x\right) g\left( x+\mu \right) \right] 
\end{equation}
as a weight function to select an ensemble of $g$ fields. The normal
gauge-fixing procedure is formally regained in the limit $\alpha \rightarrow
\infty $. Computationally, this generalized gauge-fixing procedure
 can be implemented as a Monte Carlo
simulation inside a Monte Carlo simulation. Note that the $g$ fields should
be thought of as quenched variables, since they do not effect the 
$U$-ensemble.

The expectation value of an observable $O$, gauge-invariant or not, is now
given by 
\begin{equation}
  \left\langle O\right\rangle =\frac{1}{Z_{U}}\int \left[ dU\right]
  \,e^{S_{U}\left[ U\right] }\frac{1}{Z_{gf}\lbrack U\rbrack }\int \left[
    dg\right] e^{S_{gf}\left[ U,g\right] }\,O\,, 
\end{equation}
where 
\begin{equation}
  Z_{gf}\lbrack U\rbrack =\int \left[ dg\right] e^{S_{gf}\left[ U,g\right] }
. 
\end{equation}

Formally, $g$ is just a quenched scalar field. It has two
independent symmetry groups, $G_{local}\otimes G_{global}$, so that it
appears to be in the adjoint representation of the gauge group, but the left
and right symmetries are distinct. The generating functional $Z_{gf}\lbrack
U\rbrack $ is a lattice analog of the inverse of the Fadeev-Popov
determinant \cite{Bock:2000cd}, but there are some important
differences. Note immediately that $Z_{gf}\lbrack U\rbrack $ depends on the
gauge-fixing parameter $\alpha $. More fundamentally, with the continuum
Fadeev-Popov determinant there is the vexing question of Gribov copies:
what should be done about field configurations on the same gauge orbit
satisfying the same gauge condition? The lattice formalism avoids this
question. By construction, gauge-invariant observables are evaluated by
integrating over all configurations. Gauge-variant quantities receive
weighted contributions from Gribov copies. Thus the connection between
lattice gauge fixing and gauge fixing in the continuum is not simple.

There is an apparent conflict between lattice gauge fixing and
Elitzur's theorem \cite{Elitzur:im},
which tells us that a lattice gauge symmetry cannot 
be broken, neither spontaneously nor by an explicit symmetry-breaking
term in the action. However, gauge invariance  implies that the 
lattice gluon propagator
is identically zero. 
Lattice gauge fixing avoids Elitzur's theorem by creating a new global
symmetry on top of the underlying gauge symmetry. The new global symmetry
introduced by the lattice gauge-fixing procedure is used to construct
a proxy for the gauge field with many of the same properties, but transforming
as the adjoint of the global symmetry rather than the local one.

\section{Formalism for Quenching}\label{sec:qu_form}

\begin{figure} 
\includegraphics[width=3in]{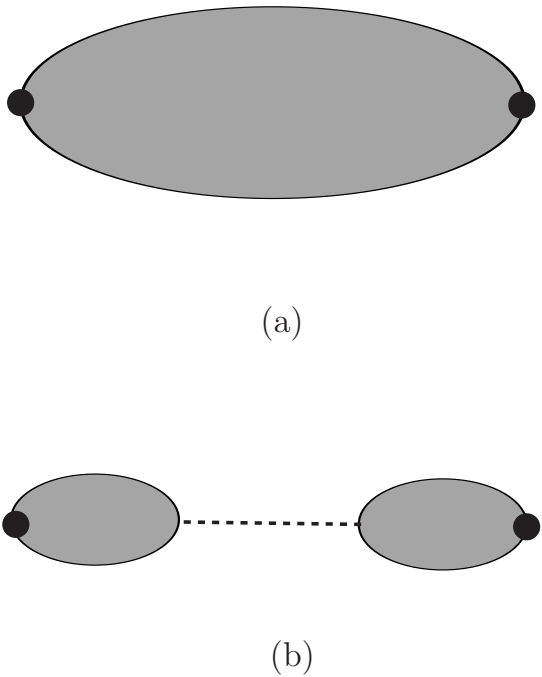}
  \caption
  	{Diagrams contributing to 
  		$\left\langle OO\right\rangle$.
  	 (a) represents the sum of all
	  $\phi_2$-connected graphs contributing to the two-point function.
	  (b) is an example of a
	  $\phi_2$-disconnected graph, where 
	  the dashed line is a $\phi_1$ propagator. }
\label{fig:phi2_graphs}
\end{figure}

In this section we develop a general formalism for a set of
quenched fields, collectively designated $\phi _{2}$ and a set of
unquenched fields, collectively designated $\phi _{1}$. These fields can be
scalars, spin-$1/2$ fermions, \textit{et cetera}. An observable
$O$ is coupled to an external source $K$. Using a
compact functional notation, the generating functional for $O$ is 
\begin{equation}
  Z\left[ K\right] =\int \left[ d\phi _{1}\right] e^{-S_{1}\left[ \phi
      _{1}\right] }\frac{\int \left[ d\phi _{2}\right] e^{-S_{2}\left[
      \phi _{2};\phi _{1}\right] +\int KO}}{\int \left[
      d\widetilde{\phi }_{2}\right] e^{-S_{2}\left[ \widetilde{\phi
      }_{2};\phi _{1}\right] }}\,.
\end{equation}
We define the generator of $\phi _{2}$-connected subgraphs 
$W\left[ K;\phi_{1}\right] $ by 
\begin{equation}
  e^{W\left[ K;\phi _{1}\right] }=\int \left[ d\phi _{2}\right]
  e^{-S_{2}\left[ \phi _{2};\phi _{1}\right] +\int KO} 
\end{equation}
so that 
\begin{equation}
  Z\left[ K\right] =\int \left[ d\phi _{1}\right] e^{-S_{1}\left[ \phi
      _{1}\right] }e^{W\left[ K;\phi _{1}\right] -W\left[ 0;\phi _{1}\right] }\,, 
\end{equation}
where the term $W\left[ 0;\phi _{1}\right]$ comes from the integration
over $\widetilde{\phi }_{2}$. 
The difference between the
quenched and unquenched models is that the quenched model has an
additional factor of $\exp \left( -W\left[ 0;\phi _{1}\right] \right) $.
This factor cancels all the internal $\phi _{2}$
loops, which are absent in the quenched model. 
The two-point function $\left\langle
OO\right\rangle $ is given by 
\begin{equation}
\left\langle OO\right\rangle =\int \left[ d\phi _{1}\right] e^{-S_{1}\left[
\phi _{1}\right] }\left[ \left( \frac{\delta ^{2}W}{\delta K^{2}}\right)
+\left( \frac{\delta W}{\delta K}\right) ^{2}\right] _{K=0}\,. 
\end{equation}
Note that there are two
contributions to $\left\langle OO\right\rangle $. The first, from $\delta
^{2}W/\delta K^{2}$, represents all the graphs which are $\phi _{2}$
-connected. A graph is $\phi _{2}$-connected if it remains connected after
cutting any number of $\phi _{1}$ lines. The second contribution, from $
\left( \delta W/\delta K\right) ^{2}$, represents $\phi _{2}$-disconnected
graphs. See Fig.~\ref{fig:phi2_graphs} for examples.

We can apply the general prescription for quenching  to gauge
theories. It is important to realize that
this analysis is aimed solely at understanding the gauge-fixing
procedure, and does not address the underlying mechanisms for mass generation.
For simplicity, we use the language of continuum field theories,
but the general form of the generating functional also holds on the lattice.
Our auxiliary fields will be $g(x)$ and its partner $\widetilde{g}(x)$ which
take on values in the gauge group $G.$ The Euclidean action for $g$ is
simply 
\begin{equation}
S_{g}=\int F^{2}Tr\left( D_{\mu }g\right) ^{+}\left( D_{\mu }g\right)\,, 
\end{equation}
where the covariant derivative is $D_{\mu }g=\partial _{\mu }g+iA_{\mu }g$.
The constant $F^{2}$ can be identified with $\alpha /2N$ in the lattice
theory. The symmetry group is $G_{local}\otimes G_{global}$, where the gauge
symmetry acts on the left, and the global symmetry acts on the right. The
proxies for the gauge fields are the conserved currents associated with the
global symmetry: 
\begin{eqnarray}
J_{\mu }^{a} &=&\frac{i}{2}Tr\left[ gT^{a}\left( D_{\mu }g\right)
^{+}-\left( Dg\right) T^{a}g^{+}\right]  \\
&=&Tr\left[ T^{a}g^{+}A_{\mu }g-\frac{i}{2}T^{a}g^{+}\overleftrightarrow{
\partial }_{\mu }g\right] .
\end{eqnarray}
If $g$ is expanded around the identity, we see the natural
identification of $J_{\mu }^{a}$ as a gauge-fixed form of $A_{\mu }^{a}$.

If we couple a source $K_{\mu }^{a}$ to the currents, then the generating
functional is 
\begin{equation}
Z\left[ K\right] =\int \left[ dA\right] e^{-S_{A}}\frac{\int \left[
dg\right] e^{-S_{g}+\int KJ}}{\int \left[ d\widetilde{g}\right] e^{-S_{
\widetilde{g}}}} 
\end{equation}
and as before, we define the generator of $g$-connected graphs to be 
\begin{equation}
e^{W\left[ K;A\right] }=\int \left[ dg\right] e^{-S_{g}+\int KJ}. 
\end{equation}

The action $S_{g}$ can be written in the form 
\begin{equation}
S_{g}=\int F^{2}Tr\left[ \left( \partial _{\mu }g\right) ^{+}\left( \partial
_{\mu }g\right) +j_{\mu }A_{\mu }+A_{\mu }^{2}\right] 
\end{equation}
where we have defined the gauge-variant currents 
\begin{equation}
j_{\mu }=i\left( \partial _{\mu }g^{+}\right) g-ig^{+}\partial _{\mu }g\text{
.} 
\end{equation}
It is important to note the distinction between $J_{\mu }$, a set of
currents transforming non-trivially under the global symmetry, and the
gauge-variant currents $j_{\mu }$, which transform under the local symmetry,
but are invariant under the global symmetry.
The generator of $g$-connected graphs is 
\begin{equation}\label{eq:eq_before_proxy_prop}
e^{W\left[ K;A\right] }
=\int \left[ dg\right] \exp \left[ -\int F^{2}Tr\left[ \left( \partial
_{\mu }g\right) ^{+}\left( \partial _{\mu }g\right) +j_{\mu }A_{\mu }+A_{\mu
}^{2}-K_{\mu }J_{\mu }\right] \right]\,. 
\end{equation}
Now we have the expression for the proxy of the gauge field propagator
\begin{equation}\label{eq:proxy_prop}
\left\langle J_{\mu }^{a}J_{\nu }^{b}\right\rangle =
\frac{1}{Z_A}
\int \left[ dA\right]
e^{-S_{A}}\left[ \left( \frac{\delta ^{2}W}{\delta K_{\mu }^{a}\delta K_{\nu
}^{b}}\right) +\left( \frac{\delta W}{\delta K_{\mu }^{a}}\right) \left( 
\frac{\delta W}{\delta K_{\nu }^{b}}\right) \right] _{K=0}\,,
\end{equation}
where 
\begin{equation}
\left( \frac{\delta W}{\delta K_{\mu }^{a}}\right) _{K=0}=\frac{\int \left[
dg\right] J_{\mu }^{a}\exp \left[ -\int F^{2}Tr\left[ \left( \partial _{\mu
}g\right) ^{+}\left( \partial _{\mu }g\right) +j_{\mu }A_{\mu }\right]
\right] }{\int \left[ dg\right] \exp \left[ -\int F^{2}Tr\left[ \left(
\partial _{\mu }g\right) ^{+}\left( \partial _{\mu }g\right) +j_{\mu }A_{\mu
}\right] \right] }
\end{equation}
gives rise to
the graphically $g$-disconnected graphs.
Similarly, the $g$-connected graphs are obtained from
\begin{equation}
\left( \frac{\delta ^{2}W}{\delta K_{\mu }^{a}\delta K_{\nu }^{b}}\right) =
\frac{\int \left[ dg\right] J_{\mu }^{a}J_{\nu }^{b}\exp \left[ -\int
F^{2}Tr\left[ \left( \partial _{\mu }g\right) ^{+}\left( \partial _{\mu
}g\right) +j_{\mu }A_{\mu }\right] \right] }{\int \left[ dg\right] \exp
\left[ -\int F^{2}Tr\left[ \left( \partial _{\mu }g\right) ^{+}\left(
\partial _{\mu }g\right) +j_{\mu }A_{\mu }\right] \right] }\,,
\end{equation}
so the total propagator consists of two terms.
Note that the apparent mass term for $A_{\mu }$ has been cancelled out.

In the case of QED, we can exactly solve for $\left\langle J_{\mu }J_{\nu
}\right\rangle $. Writing $g$ as $\exp \left( i\theta \right) $, we have $
S_{g}=\int F^{2}\left( \partial _{\mu }\theta +A_{\mu }\right) ^{2}$ and the
current $J_{\mu }$ is given by $J_{\mu }=A_{\mu }+\partial _{\mu }\theta $.
The generating functional is 
\begin{equation}
e^{W\left[ K;A\right] }=\int \left[ d\theta \right] \exp \left[ -\int \frac{
v^{2}}{2}\left( \partial _{\mu }\theta +A_{\mu }\right) ^{2}+\int K_{\mu
}\left( \partial _{\mu }\theta +A_{\mu }\right) \right] 
\end{equation}
where $v^{2}=2F^{2}$. Then 
\begin{eqnarray}
	e^{W\left[ K;A\right] }
	&=&\exp \left[ -\int \frac{v^{2}}{2}A_{\mu }^{2}+\int K_{\mu }A_{\mu
	}\right] \det \left( -v^{2}\partial ^{2}\right) ^{-1/2} \nonumber\\
	&& \times \exp \left[ \frac{1}{2
	}\int \left( K_{\mu }-v^{2}A_{\mu }\right) \frac{1}{v^{2}}\frac{\partial
	_{\mu }\partial _{\nu }}{\partial ^{2}}\left( K_{\nu }-v^{2}A_{\nu }\right)
	\right]\,.
\end{eqnarray}
The momentum space form of the propagator is 
\begin{equation}
\left\langle J_{\mu }\left( k\right) J_{\nu }\left( -k\right) \right\rangle
=\left( \delta _{\mu \rho }-\frac{k_{\mu }k_{\rho }}{k^{2}}\right)
\left\langle A_{\rho }\left( k\right) A_{\sigma }\left( -k\right)
\right\rangle \left( \delta _{\sigma \nu }-\frac{k_{\sigma }k_{\nu }}{k^{2}}
\right) +\frac{1}{v^{2}}\frac{k_{\mu }k_{\nu }}{k^{2}}\,. 
\end{equation}
The first term is the photon propagator, projected onto the transverse
subspace, and the second term represents a direct contribution from the $
\theta $ field. Note that as $v$ goes to infinity, the propagator becomes
purely transverse.

It is amusing to note that the above formula is also valid in the
unquenched case, where no $\widetilde{\theta }$ field is introduced.
The essential change is that the gauge field acquires a mass $v$ via
the simplest example of the Higgs mechanism. The propagator for
$\partial _{\mu }\theta +A_{\mu }$ has exactly the same form as in
the quenched case, but the mass term now causes the $A_{\mu }$
two-point function to have the explicit form
\begin{equation}
  \left\langle A_{\rho }\left( k\right) A_{\sigma }\left( -k\right)
  \right\rangle = \frac{\delta _{\rho \sigma }+\frac{k_{\rho
      }k_{\sigma }}{v^{2}}}{k^{2}+v^{2}}
\end{equation}
and one easily checks that 
$\left\langle J_{\mu }\left( k\right) J_{\nu }\left( -k\right) \right\rangle
=\left\langle A_{\mu }\left( k\right) A_{\nu }\left( -k\right) \right\rangle$
in the
unquenched case.

Lattice simulations of the $U(1)$ propagator \cite{Coddington:yz}, 
which correspond to the limit $v\rightarrow \infty $, give a
non-zero asymptotic mass in the strong-coupling, confining region, and a
zero mass in the weak-coupling, free field region. In neither region are
violations of spectral positivity observed, consistent with our results here.

\section{Soluble Examples: Mixing Models}\label{sec:mix_models}

In our previous work \cite{Aubin:2003ih}, we analyzed the simplest
model displaying spectral positivity violation due to quenching: a
model of two free, real scalar fields with a non-diagonal mass matrix,
with one of the fields quenched. Here we generalize this to include
the effects of mixing in the kinetic terms as well. The Lagrangrian
for the scalar mixing model is
\begin{equation}
L=\frac{1}{2}\left[ \left( \partial \phi _{1}\right) ^{2}+m_{1}^{2}\phi
_{1}^{2}\right] +\frac{1}{2}\left[ \left( \partial \phi _{2}\right)
^{2}+m_{2}^{2}\phi _{2}^{2}\right] -\mu ^{2}\phi _{1}\phi _{2}-\varepsilon
\left( \partial \phi _{1}\right) \left( \partial \phi _{2}\right) 
\end{equation}
where we take $\phi _{2}$ to be quenched. If $\phi _{2}$ were not quenched,
this Lagrangian could be rewritten in terms of two free massive fields after
a field redefinition. The quenched model can be solved by the functional
method described above. We have done this in the case of $\varepsilon =0$ in
Ref.~\cite{Aubin:2003ih}. It suffices here to note that the
Dyson series for the $\phi _{2}$ propagator, truncated by quenching, is given by
\begin{equation}
\frac{1}{p^{2}+m_{2}^{2}}+\frac{1}{p^{2}+m_{2}^{2}}\left( \mu
^{2}+\varepsilon p^{2}\right) \frac{1}{p^{2}+m_{1}^{2}}\left( \mu
^{2}+\varepsilon p^{2}\right) \frac{1}{p^{2}+m_{2}^{2}}\,. 
\end{equation}
The $\phi _{2}$ propagator can also be written in the form

\begin{equation}
\allowbreak \frac{A}{p^{2}+m_{2}^{2}}+\allowbreak \frac{B}{\left(
p^{2}+m_{2}^{2}\right) ^{2}}+\frac{C}{p^{2}+m_{1}^{2}}\allowbreak 
\end{equation}
where $A$, $B$, and $C$ are functions of the parameters of the
Lagrangian.

Spectral positivity is violated, because a double pole is a limiting case of
a negative-metric contribution: 
\begin{equation}
\frac{1}{\left( p^{2}+m_{2}^{2}\right) ^{2}}=\lim_{m\rightarrow m_{2}}\frac{1
}{m_{2}^{2}-m^{2}}\left[ \frac{1}{p^{2}+m^{2}}-\frac{1}{p^{2}+m_{2}^{2}}
\right] . 
\end{equation}
The form in coordinate space is very interesting. In any number of
dimensions, we can consider propagators using wall sources, \textit{i.e.},
sources of co-dimension $1$. This sets the momentum equal to
zero in all the directions except the direction perpendicular to the wall.
The coordinate space form of the two-point function is
\begin{equation}
\frac{A}{2m_{2}}e^{-m_{2}\left| x\right| }+\frac{B}{4m_{2}^{3}}
e^{-m_{2}\left| x\right| }\left( 1+m_{2}\left| x\right| \right) +\frac{C}{
2m_{1}}e^{-m_{1}\left| x\right| }. 
\end{equation}
It is the second term, arising from the
double pole, which violates spectral positivity.

\begin{figure} 
\includegraphics[width=4in]{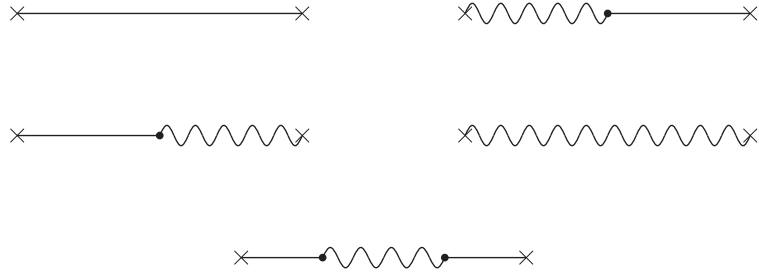}
  \caption{Diagrams that contribute to the propagator in 
  the vector mass-mixing model. Wavy lines represent the $A_\mu$ propagator,
  and straight lines the $B_\mu$ propagator. }
\label{fig:vector_mix}
\end{figure}

Propagators are more complicated when the fields couple to both
quenched and unquenched states. 
Consider a model of the gauge-fixed vector field
as a mixture of two massive vector fields: $B_{\mu }^{a}$, which will be
quenched, and $A_{\mu }^{a}$, which is unquenched. The Lagrangian density can be taken
to be
\begin{equation}
L=\frac{1}{4}\left( \partial _{\mu }A_{\nu }^{a}-\partial _{\nu }A_{\mu
}^{a}\right) ^{2}+\frac{1}{4}\left( \partial _{\mu }B_{\nu }^{a}-\partial
_{\nu }B_{\mu }^{a}\right) ^{2}+\frac{1}{2}m_{1}^{2}\left( A_{\mu
}^{a}\right) ^{2}+\frac{1}{2}m_{2}^{2}\left( B_{\mu }^{a}\right) ^{2}+\mu
^{2}A_{\mu }^{a}B_{\mu }^{a} 
\end{equation}
with the sum over indices implicit. If both fields were unquenched, then the
mass eigenstates would be 
\begin{equation}\label{eq:mass_eigen}
\frac{m_{1}^{2}+m_{2}^{2}\pm \sqrt{\left( m_{1}^{2}-m_{2}^{2}\right)
^{2}+4\mu ^{4}}}{2}. 
\end{equation}
Consider a field which is a linear combination of $A$ and $B$, given by $
\kappa A+\lambda B$. Only the relative sign of $\kappa $ and $\lambda $ is
important. The use of wall sources greatly simplifies what would be a complicated
tensor structure, and the relevant Dyson series for the transverse components
of the propagator is 
\begin{equation}\label{eq:vec_mix_prop}
\frac{\kappa ^{2}}{p^{2}+m_{1}^{2}}+\frac{\lambda ^{2}}{p^{2}+m_{2}^{2}}-2
\frac{\kappa \lambda \mu ^{2}}{\left( p^{2}+m_{1}^{2}\right) \left(
p^{2}+m_{2}^{2}\right) }+\frac{\lambda ^{2}\mu ^{4}}{\left(
p^{2}+m_{1}^{2}\right) \left( p^{2}+m_{2}^{2}\right) ^{2}} 
\end{equation}
corresponding to the diagrams shown in Fig.~\ref{fig:vector_mix}.

The propagator can be rearranged in several forms. The factored form 
\begin{equation}
\label{eq:factored_prop}
\frac{\lambda ^{2}}{p^{2}+m_{2}^{2}}+\left( 1-\frac{\lambda \mu ^{2}/\kappa 
}{\left( p^{2}+m_{2}^{2}\right) }\right) \frac{\kappa ^{2}}{p^{2}+m_{1}^{2}}
\left( 1-\frac{\lambda \mu ^{2}/\kappa }{\left( p^{2}+m_{2}^{2}\right) }
\right) 
\end{equation}
is helpful because it shows the beginning of the infinite Dyson series in
the case where both fields are unquenched.
The partial fraction form explicitly presents all
the information available: there is a single pole in $p^{2}$ at $-m_{1}^{2}$,
a single and a double pole at $p^{2}=-m_{2}^{2}$, and three coefficients
giving the residues at the poles. 
\begin{eqnarray}
&&\left( \kappa +\frac{\lambda \mu ^{2}}{m_{1}^{2}-m_{2}^{2}}\right) ^{2}
\frac{1}{p^{2}+m_{1}^{2}} \nonumber\\
&&+\left[ -\left( \frac{\lambda \mu ^{2}}{\left( m_{1}^{2}-m_{2}^{2}\right) }
+\kappa \right) ^{2}+\lambda ^{2}+\kappa ^{2}\right] \frac{1}{p^{2}+m_{2}^{2}
} \nonumber\\
&&+\frac{\lambda ^{2}\mu ^{4}}{\left( m_{1}^{2}-m_{2}^{2}\right) }\frac{1}{
\left( p^{2}+m_{2}^{2}\right) ^{2}}\,.
\end{eqnarray}
Note that the residue of the $m_{1}$ pole must be positive, but this
need not be true for the other residues, which come from the quenched fields.

\section{Phase Diagram of the Lattice Model}\label{sec:phase_diag}

The unquenched form of the lattice model is a Higgs model, of a type first
analyzed by Fradkin and Shenker \cite{Fradkin:1978dv}.
It is very useful to recall their analysis.
For $\alpha $ and $\beta $ small, there is a convergent strong-coupling
expansion associated with a phase in which the $g$ fields are confined into
bound states. For $\alpha $ and $\beta $ large, perturbation theory
indicates that the Higgs mechanism takes place in its most complete form,
with no remaining scalar fields. At tree level, the only particles are
massive vector particles. Naively, there appears to be two distinct phases,
a confined phase and a Higgs phase. However, the field $g$ is in the
fundamental representation of the gauge group, and breaks the $Z(N)$
symmetry associated with confinement in the pure gauge theory, in a manner
similar to quarks. As Fradkin and Shenker showed, there is no absolute
distinction between the confined and Higgs phases in this case, and the two
regions of the lattice phase diagram are connected. 

\begin{figure} 
\includegraphics[width=5in]{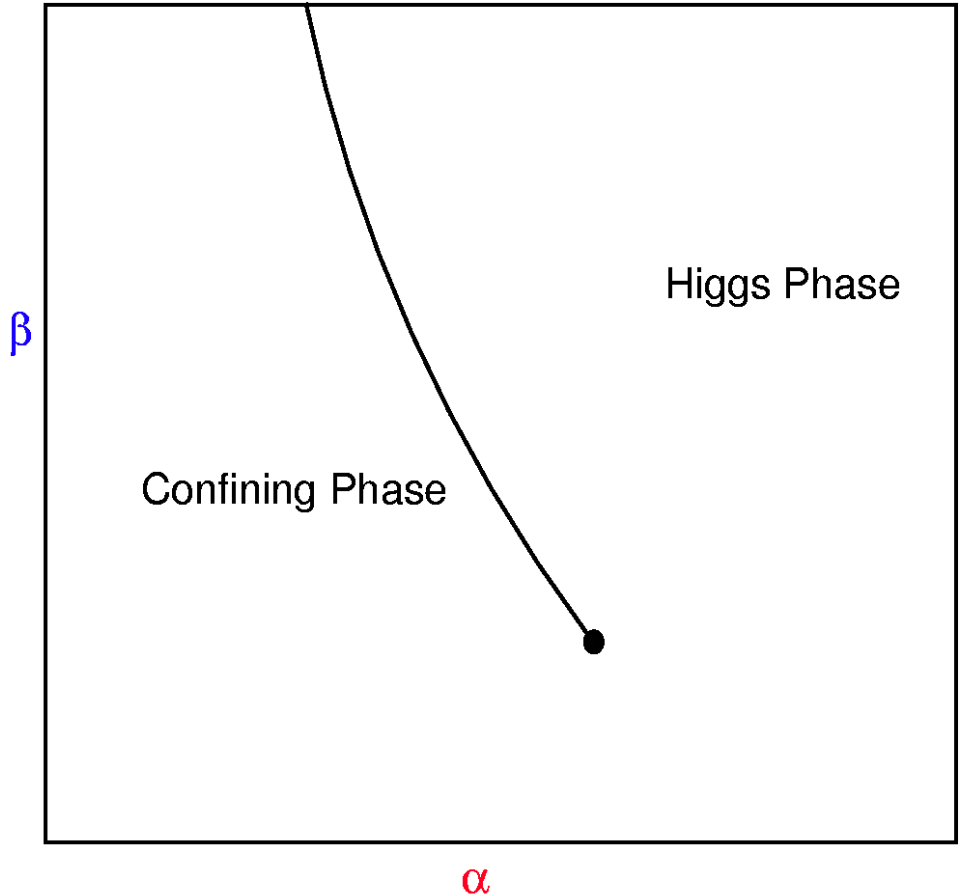}
  \caption{Phase diagram in the $\alpha-\beta$ plane for the $SU(2)$ gauge
  theory.}
\label{fig:beta_alpha}
\end{figure}

\begin{figure} 
\includegraphics[width=4in, angle=270]{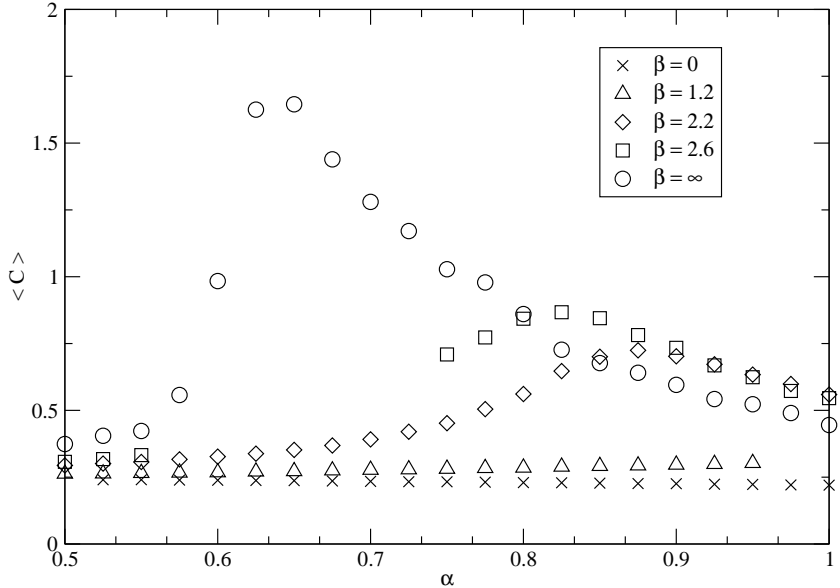}
  \caption{Plot of the specific heat for the quenched
  model as a function of $\alpha$.}
\label{fig:jump}
\end{figure}

We have studied the phase diagram of the quenched model using lattice
simulations, and the results are similar to those for the unquenched
model. As shown schematically in Fig.~\ref{fig:beta_alpha}, there is
a critical line coming out of $\beta =\infty $. The origin of the line
is the same in both the quenched and unquenched models: at $\beta
=\infty $, the link fields become gauge transforms of the identity,
and can be absorbed into the $g$ fields. Thus setting $\beta =\infty $
gives a spin model with a phase transition between a disordered phase
for small $\alpha $ and an ordered phase for large $ \alpha $. As
$\beta $ decreases, the disordering effect of the gauge fields
increases, requiring larger $\alpha $ for the phase transition. In
both the quenched and unquenched cases, the critical line does not
completely separate the two phases, but terminates in a critical end
point in the $\alpha -\beta $ plane. This line appears to be first
order in the quenched case, as is the case in the unquenched
model. Figure \ref{fig:jump} shows the specific heat for the quenched
model as $\alpha $ is varied for various values of $\beta $ in the
case of $SU(2)$.

As first discussed by Fradkin and Shenker in the unquenched case, the
interpretation of the massive vector particles of the model depends on which
part of the phase diagram is being considered. In the confined phase, the
vector masses are associated with vector bound states of the confined
scalars. This interpretation is the same in both the quenched and unquenched
models. For the unquenched model, the vector particles are interpreted as
fundamental particles, made massive via the Higgs mechanism. In the quenched
model, this interpretation is problematic. As we have seen in our continuum
treatment of gauge fixing as quenching, the term responsible for giving the
vector fields a mass cancels out, and the gauge fields do not acquire a mass
via the Higgs mechanism. The gauge fields can acquire a mass via their own
self-interactions, however.

\subsection{Interpretation as confined theory}

\begin{figure}
  \includegraphics[width=3in]{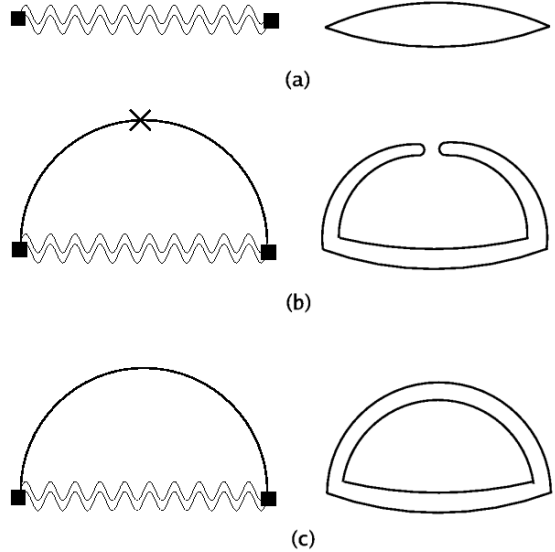}
  \caption{Diagrams contributing to the $a_0$ propagator in quenched
	  	QCD or the vector propagator in the confined phase of
	  	a gauge-fixing model.  Diagram (c) only contributes in
	  	the unquenched case. }
	  \label{fig:conf_dia}
\end{figure}

We will interpret the vector multiplet states in the confined region
in a manner similar to the work of Bardeen et al.\ on the $a_{0}$ in
quenched QCD \cite{Bardeen:2001yz,Bardeen:2001jm}. The $\eta ^{\prime }$ propagator
violates spectral positivity in quenched QCD. Within chiral
perturbation theory for full QCD, the $\eta ^{\prime }$ becomes
heavier than the particles in the pseudoscalar multiplet by the
summation of a mass insertion term arising from the anomaly.  In
quenched QCD, the summation of the Dyson series is truncated, giving
rise to a double pole in the $\eta ^{\prime }$ propagator.  The
effects of this show up in the $a_{0}$ propagator via loops containing
an $\eta ^{\prime }$. The left-hand side of Fig.~\ref{fig:conf_dia}(a)
shows the contribution of a single $a_{0}$ particle to the $a_{0}$
propagator, and the right-hand side shows an equivalent diagram in
terms of quarks.  Figures~\ref{fig:conf_dia}(b) and (c) show loop
contributions from states containing an $\eta ^{\prime }$ as part of
the intermediate state. Figure~\ref{fig:conf_dia}(b) differs from
Fig.~\ref{fig:conf_dia}(c) by a single mass insertion. In full QCD,
these are the first two terms in a geometric series which can be
easily summed to give a heavy $\eta ^{\prime }$.  In quenched QCD,
Fig.~\ref{fig:conf_dia}(c) does not contribute, because it has
an internal quark loop. As shown by Bardeen et al., the
bubble in Fig.~\ref{fig:conf_dia}(b) leads to spectral positivity violation in the $a_{0}$
propagator.

We can take over this argument to case of gauge fixing, where the
gauge-fixed vector operator must be
understood as creating bound states in the confined region. In
this extended analogy, there is a $g^{+}Ag$ vector bound state,
whose propagation is represented by Fig.~\ref{fig:conf_dia}(a).
There is also a contribution from intermediate states containing
a isoscalar scalar state, as in Fig.~\ref{fig:conf_dia}(b).
The isoscalar scalar
thus plays a role in the confining region of the 
quenched theory similar to that of the $\eta ^{\prime }$
in quenched QCD.

Let $B(p)$ stand for the bubble diagram and let the vector
propagator be given by $V(p)$. We parametrize the coupling between $V$ and $
B $ as $-\mu ^{2}$. Suppose we look at the propagator for an operator that
couples with strength $\kappa $ to the bubble and strength $\lambda $ to the
vector. Then the propagator can be written in the form 
\begin{equation}
\lambda ^{2}V(p)+\left( 1-\frac{\lambda \mu ^{2}}{\kappa }V(p)\right) \frac{
\kappa ^{2}B(p)}{1+\mu ^{2}B(p)V(p)}\left( 1-\frac{\lambda \mu ^{2}}{\kappa }
V(p)\right) . 
\end{equation}
We have chosen our parameters, including the sign of $\mu ^{2}$, \ to
facilitate comparison with \eq{factored_prop}. If we identify the pole in the
vector propagator $V(p)$ with $-m_{2}^{2}$, and the pole in
the resummed bubble $\kappa ^{2}B(p)/\left( 1+\mu ^{2}B(p)V(p)\right) $ with 
$-m_{1}^{2}$, then on shell we have exactly reproduced \eq{factored_prop}.

\subsection{Interpretation in Higgs region}

In order to discuss the region where $\alpha $ is large, we return to the
continuum model of Sec.~\ref{sec:qu_form}, and write $g\left( x\right) $ as 
$\exp \left( iH(x)/F \right)$ where $H$ is Hermitian and traceless.
We assume that for $F$ sufficiently large, we are in the Higgs phase,
where $g\left( x\right) $ can be expanded around the identity.
We are only treating $g\left( x\right) $ perturbatively,
and non-perturbative phenomena in the gauge-field sector,
such as dynamical mass generation, are not treated here.

In the unquenched theory, the Higgs mechanism
would occur via the term in $S_g$
\begin{equation}
\int F^{2}Tr\left( A_{\mu }g\right) ^{+}\left( A_{\mu }g\right) =\int
F^{2}TrA_{\mu }^{2}\,, 
\end{equation}
but this term is explicitly canceled in the quenched theory. Internal loops
associated with $H$ are also cancelled out in the quenched theory, as
discussed in Sec.~\ref{sec:qu_form}. 
The
gauge-invariant current $J_{\mu }^{a}$, which is a proxy for the gauge field,
is 
\begin{eqnarray}
	J_{\mu }^{a} &=&Tr\left[ T^{a}g^{+}A_{\mu }g-\frac{i}{2}T^{a}g^{+}
	\overleftrightarrow{\partial }_{\mu }g\right] \nonumber\\
	&=&Tr\Biggl[ T^{a}\biggl( A_{\mu }-\frac{i}{F}\left[ H,A_{\mu }\right] 
	+\frac{1}{F}\partial _{\mu }H-\frac{1}{2F^{2}}\left[ H,\left[ H,A_{\mu }\right]
	\right] \nonumber\\*
		&&{}-\frac{i}{2F^{2}}\left[ H,\partial _{\mu }H\right] \biggr)
	+O(H^{3})\Biggr]\,,
\end{eqnarray}
so the current $J_{\mu }^{a}$ has a term with $A_{\mu }^{a}$ alone. On the
other hand, the current which couples perturbatively to $A_{\mu }$ is 
\begin{eqnarray}\label{eq:current_higgs}
  j_{\mu } &=&i\left( \partial _{\mu }g^{+}\right) g-ig^{+}
  \partial _{\mu }g \nonumber\\
  &=&\frac{2}{F}\partial _{\mu }H+\frac{i}{F^{2}}\left[ \partial _{\mu
    }H,H\right] +O(H^{3})
\end{eqnarray}
so that $J_{\mu }$ and $j_{\mu }$ have some common terms.

\begin{figure}
  \includegraphics[width=5in]{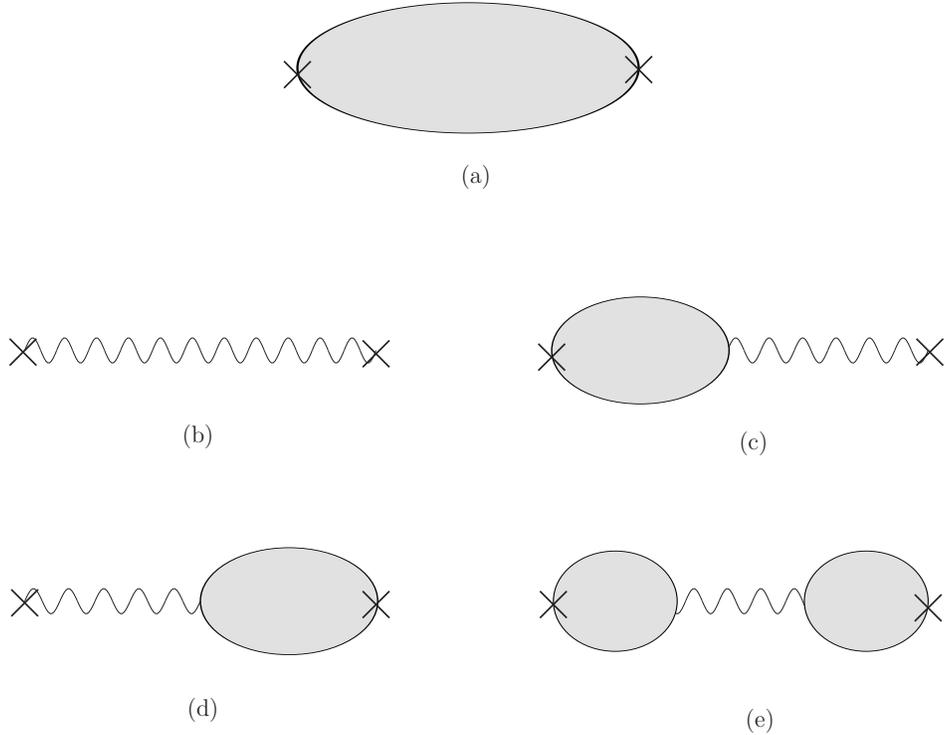}
  \caption{Schematic diagrams contributing to the vector propagator
	    in the Higgs phase. (a) consists of all $g$-connected 
	    contributions, while (b)-(e) are the $g$-disconnected 
	    pieces with single gluon exchange.}
	  \label{fig:higgs_dia}
\end{figure}

The propagator $\left\langle J_{\mu }^{a}J_{\nu }^{b}\right\rangle $ is
given by \eq{proxy_prop}
\begin{equation}
  \left\langle J_{\mu }^{a}J_{\nu }^{b}\right\rangle =
  \int \left[dA\right] e^{-S_{A}} 
  \left[ \left( \frac{\delta ^{2}W}{\delta K_{\mu}^{a}
      \delta K_{\nu }^{b}} \right) 
    +\left( \frac{\delta W}{\delta K_{\mu }^{a}}\right) 
    \left( \frac{\delta W}{\delta K_{\nu}^{b}}\right)
    \right] _{K=0}.\nonumber
\end{equation}
The first term represents $g$-connected graphs, as shown in 
Fig.~\ref{fig:higgs_dia}(a). The
second term represents all other contributions, including intermediate
states involving one or more $A_{\mu }$ propagators. If we assume that the
intermediate state with one $A_{\mu }$ propagator dominates, then we need
only consider the graphs shown in Fig.~\ref{fig:higgs_dia}(b)-(e). 
The sum of graphs has the
same structure seen in the confined phase.

Comparison of Fig. \ref{fig:higgs_dia} with \eq{factored_prop}
immediately suggests that the double pole associated with the mass $m_{2}$ can
be understood as arising from 
Fig.~\ref{fig:higgs_dia}(e). 
This does not fix the origin of the $m_1$ pole. One possibility is
that $m_1$ is associated with Fig.~\ref{fig:higgs_dia}(b),
in which case $m_1$ would be a candidate for what we mean by the term gluon
mass. As we will see in the next section, lattice simulations indicate
that the mass $m_1$ is too heavy to represent a phenomenologically
realistic gluon mass. Another possibility is that both
the $m_{1}$ and $m_{2}$ poles occur in Fig.~\ref{fig:higgs_dia}(a),
as happens in the confined region. From this point of view,
the diagrams in Fig.~\ref{fig:conf_dia}(a) and (b), which are both
$g$-connected, could move smoothly into Fig.~\ref{fig:higgs_dia}(e).
Thus the interpretation of the masses is ambiguous in this phase.

\section{Lattice Results for Propagators}\label{sec:lat_results}

We have performed simulations with the action in \eq{wilson_action}
for the case of $SU(2)$ $20^{4}$ lattices at $\beta =2.4$ and $\alpha
=\left\{ 1,1.2,1.4,1.6,1.75,2.0,2.5,3.0\right\} $. The critical line
in the $\left( \alpha ,\beta \right) $ plane lies near $\alpha =0.8$,
so most of these points are well inside the Higgs region of the phase
diagram. For each value of $\alpha $, we have performed four
independent simulations. Fits were carried out for each dataset, and
finally combined to give final results for the fitting parameters. We
begin the analysis in coordinate space to extract the masses from the
form
\begin{equation}
  G(x)=(a+bm_{2}x)e^{-m_{2}x}+ce^{-m_{1}x}+(x\rightarrow L-x)\ .
\end{equation}
In order to extract the masses, we first cut the data at $x=x_{0}$, and fit
the term $(a+bm_{2}x)e^{-m_{2}x}+(x\rightarrow L-x)$ to all points $x\geq
x_{0}$. The value of $x_{0}$ is chosen to achieve the best fit, and depends
on the dataset, but in all cases $x_{0}>1/m_{1}$. This then gives us a value
for the mass $m_{2}$ (and $a$ and $b$). Once this has been fit, we subtract
this from the original data set and then fit the remainder to 
$ce^{-m_{1}x}+(x\rightarrow L-x)$ to extract the mass $m_{1}$, as well as $c$.

The coordinate space form of the propagator is sufficient for the
determination of the two mass parameters. The coefficients $\kappa$, 
$\lambda$, and $\mu^2$ are
complicated functions of $a$, $b$ and $c$ in the coordinate space
propagator. Thus it is easiest to extract the mass parameters $m_{1}$ and 
$m_{2}$ in coordinate space, but determine the other parameters in momentum
space. After Fourier transforming the propagators, we multiply by the factor 
$(p^{2}+m_{2}^{2})^{2}(p^{2}+m_{1}^{2})$, where $p^{2}$ should be understood
as the one-dimensional lattice momentum squared, $\sin ^{2}(p)$.
We then fit 
$(p^{2}+m_{2}^{2})^{2}(p^{2}+m_{1}^{2})G(p)$ to a quadratic polynomial in
$p^{2}$. The
coefficients of the polynomical
can be extracted cleanly, and from these we can determine the
parameters from \eq{vec_mix_prop}, $\mu ^{2}$, $\kappa $ and $\lambda $.
Empirically, we
find that these parameters are only weakly dependent on the mass parameters.
Finally we have taken the values found with these fitting routines and have
averaged over the four sets of data for each value of $\alpha$, where the
error bars are determined from calculating the standard error as these data
are statistically independent. 

\begin{figure} 
\includegraphics[width=5in]{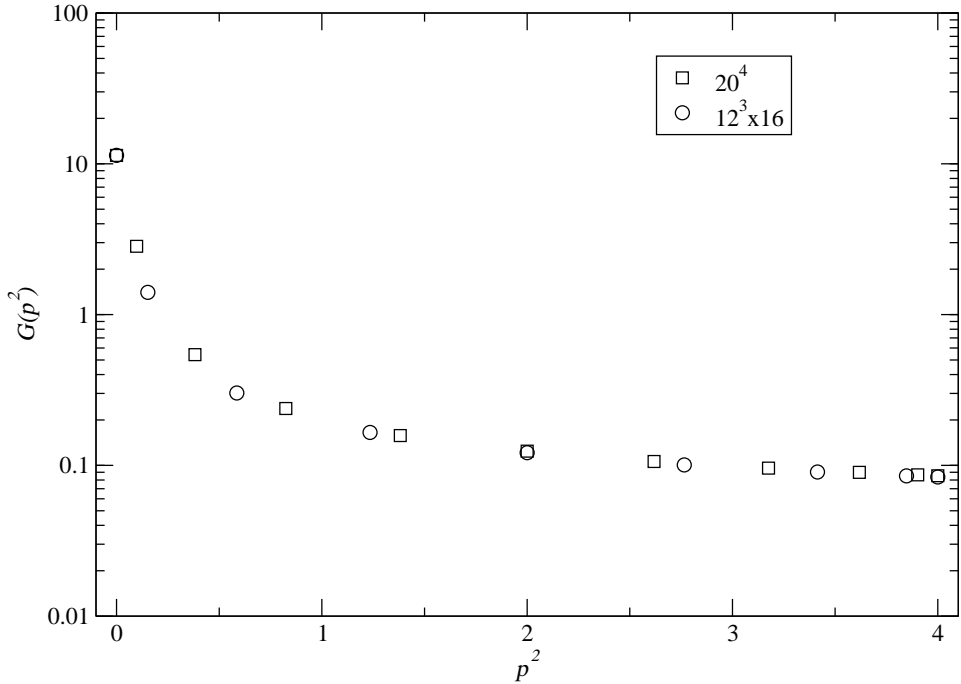}
  \caption{The momentum space gluon propagator on the $12^{3}\times
  16$ and $20^{4}$ lattices at $\alpha=1.2$.}
\label{fig:props1}
\end{figure}

\begin{figure} 
\includegraphics[width=5in]{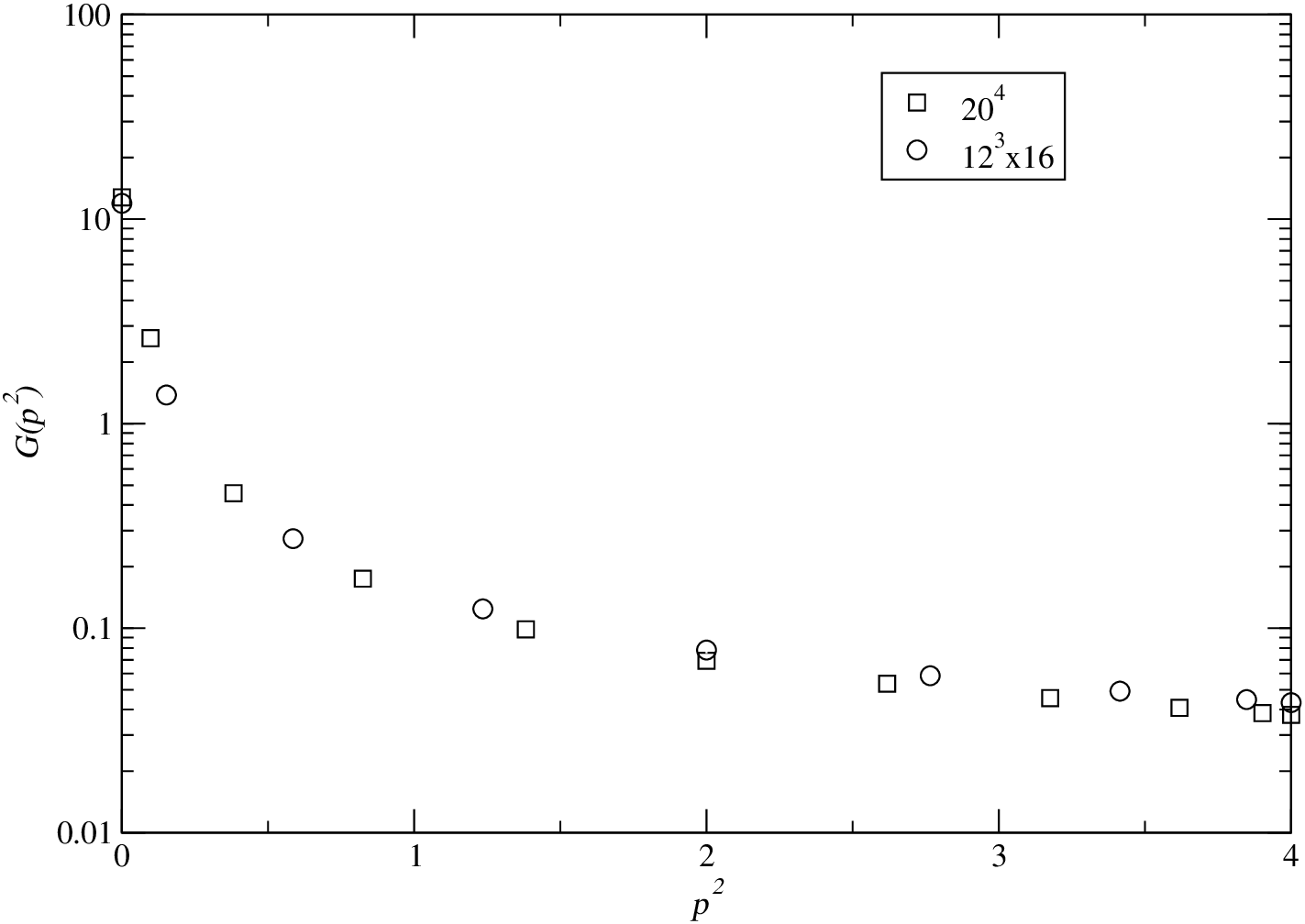}
  \caption{The momentum space gluon propagator on the $12^{3}\times
  16$ and $20^{4}$ lattices at $\alpha=2.0$.}
\label{fig:props2}
\end{figure}

In Figs.~\ref{fig:props1} and \ref{fig:props2} 
we compare our $20^{4}$ propagator results with our earlier
work \cite{Aubin:2003ih} on a $12^{3}\times 16$ lattice at $\beta = 2.4$ 
for the values $\alpha = 1.2$ and $2.0$ respectively.
Errors are smaller than the size of the data points.  
The consistency of the data and the absence of any finite-size
effects is clear, as expected from the magnitude of the masses.

In Figs.~\ref{fig:masses16} and \ref{fig:masses20} 
we plot $m_{1}$ and $m_{2}$ as a function of $\alpha $
on both $12^{3}\times 16$ and $20^{4}$ lattices. We note that $m_{2}$
appears to be essentially independent of $\alpha $ over the range studied,
and the results from the two lattice sizes are quite consistent. The mass
parameter $m_{1}$ shows definite $\alpha $ dependence for smaller values of $
\alpha $. The disagreement in the results for $m_{1}$ for the two lattice
sizes is not statistically significant, but $m_{1}$ is clearly not as
well-determined as $m_{2}$. This is to be expected, since $m_{1}$ is much
larger than $m_{2}$, and is extracted from the same propagator.

\begin{figure} 
  \includegraphics[width=5in]{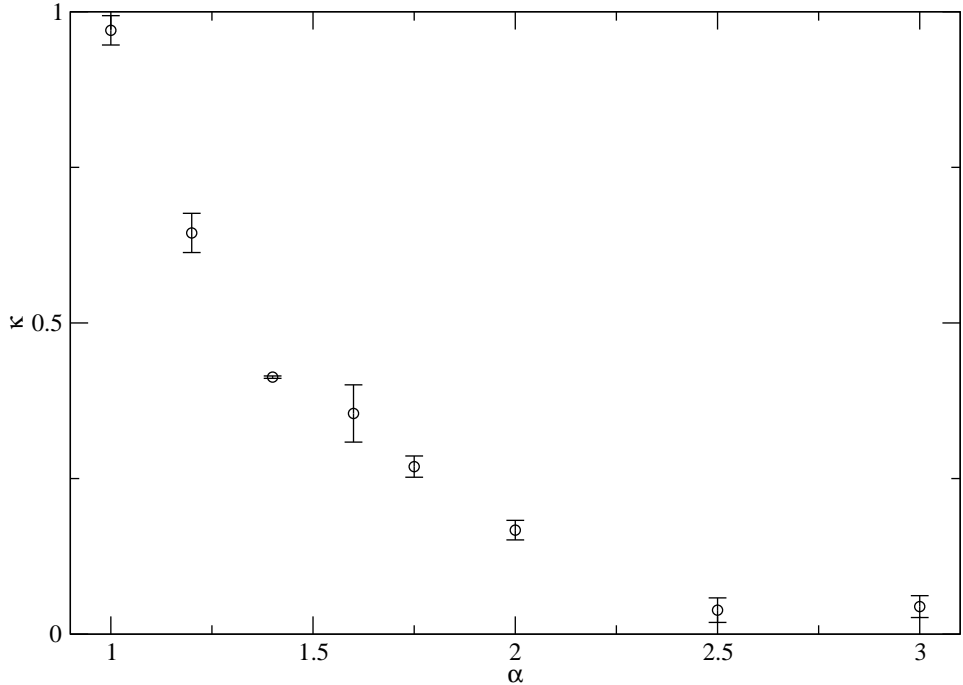}
  \caption{$\kappa$ as a function of $\alpha$.}
	  \label{fig:kappa}
\end{figure}

\begin{figure} 
  \includegraphics[width=5in]{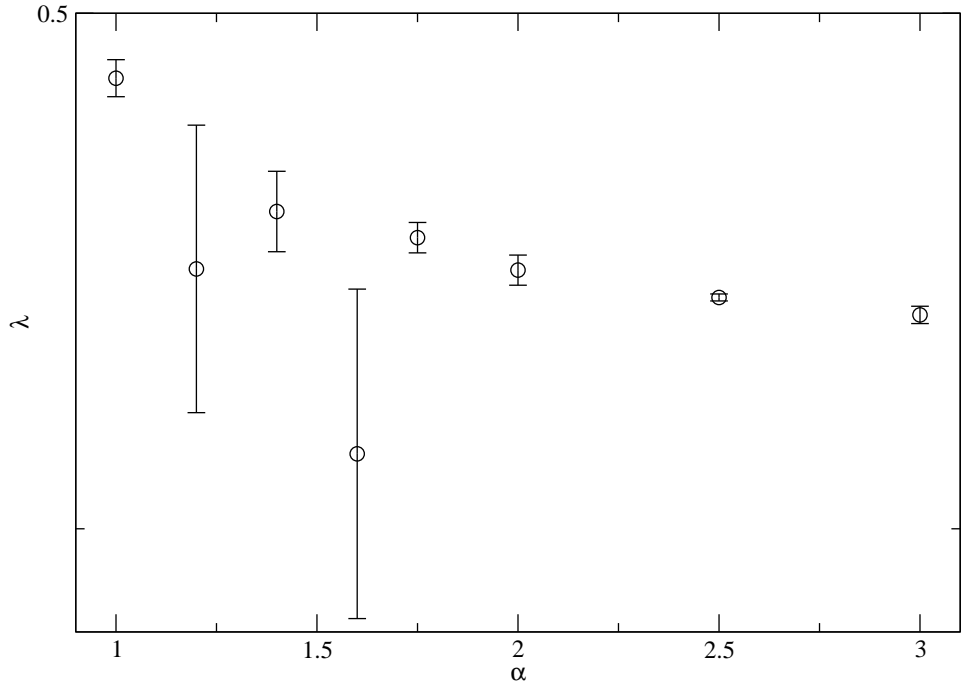}
  \caption{$\lambda$ as a function of $\alpha$.}
	  \label{fig:lambda}
\end{figure}

\begin{figure} 
  \includegraphics[width=5in]{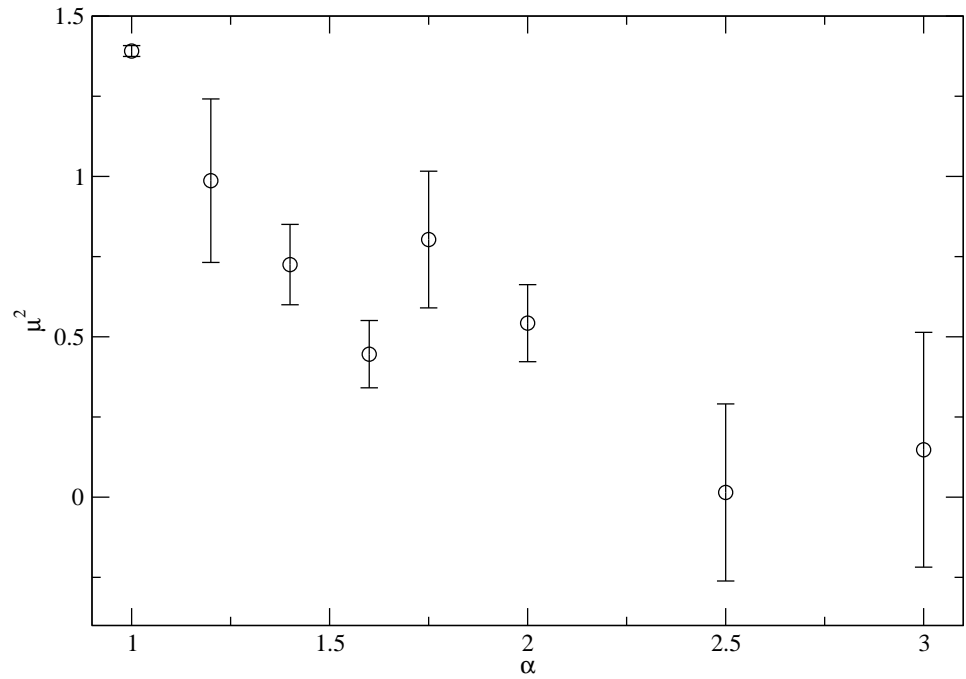}
  \caption{$\mu^2$ as a function of $\alpha$.}
	  \label{fig:mu2}
\end{figure}

The parameters $\kappa $, $\lambda $, and $\mu ^{2}$ are shown as functions
of $\alpha $ in Figs.~\ref{fig:kappa}-\ref{fig:mu2}. 
The parameter $\kappa $ represents the
coupling of the lattice gluon operator to the unquenched particle, which has
mass $m_{1}$. It is monotonically decreasing as a function of $\alpha $, but
it is difficult to determine if it asymptotes to a non-zero value for large $
\alpha $. The parameter $\lambda $ is the coupling of the gluon operator to
the quenched particle, which has mass $m_{2}$. This parameter is not as
well-determined as $\kappa $, but is much larger for large $\alpha $. The
parameter $\mu ^{2}$, which represents the mixing of the quenched and
unquenched particles, show a downward trend with increasing $\alpha $, but
with large statistical errors.

\begin{figure}
  \includegraphics[width=5in]{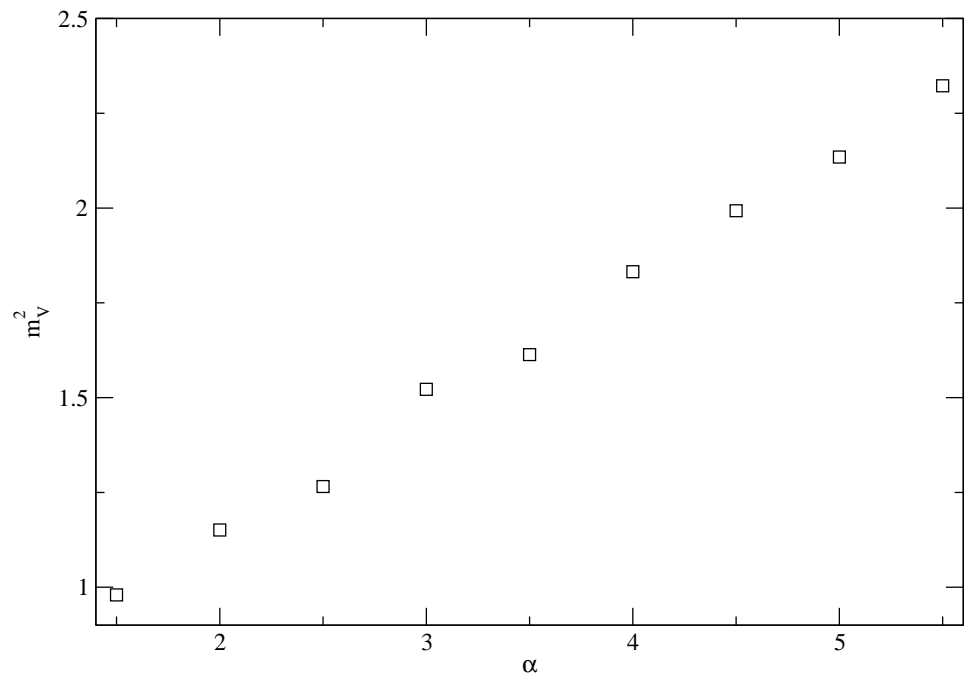}
  \caption{The mass-squared determined from fitting the 
    gluon propagator in the unquenched theory to a single pole 
    form. The linear trend is obvious.}
   \label{fig:unq_mass}
\end{figure}

The behavior of the two masses $m_{1}$ and $m_{2}$ as functions of
$\alpha $ are completely different from the behavior of the vector
particle in the unquenched model. As shown in Fig.~\ref{fig:unq_mass},
the vector mass $m_{V}$ in the unquenched model rises as $\alpha
^{1/2}$ over this range of $\alpha $, all at $\beta =2.6$. The
unquenched propagator fits very well to a single pole form in this
region. This is completely consistent with the Higgs interpretation of
the unquenched model, where $m_{V}^{2}$ is naively proportional to
$\alpha $ in physical gauge. Suppose for the moment that we were
simulating the mixing model instead of an $SU(2)$ gauge theory. Then the
parameters $m_{1}$, $m_{2}$, and $\mu $ from the quenched simulation
could be used to determine the mass eigenstates via
\eq{mass_eigen}. This procedure is dubious for the gauge theory,
because only one multiplet of vector mass eigenstates occurs in the
unquenched model, not two. Furthermore, the parameters of the quenched
and unquenched models are different due to quenching. Nevertheless, if
the parameters of the unquenched model are used in \eq{mass_eigen},
the larger of the two masses is commensurate with the single mass
observed in the unquenched model.

Our analysis of the gluon propagator reveals three parameters with
dimensions of mass: $m_{1}$, $m_{2}$, and $\mu $. As we have discussed above,
there are arguments for
both $m_{1}$ and $m_{2}$ to be interpreted as
a gluon mass parameter. The mass $m_{1}$ is the mass of the unquenched
state, and it is natural to assume that it is the gluon mass. However,
the mass of the lightest glueball  in $SU(2)$ is given by
$m_{G}/\sqrt{\sigma }=3.74\pm 0.12$ \cite{Teper:1998kw}, and 
$a\sqrt{\sigma }=0.1326$, so $m_{G}a=0.4959$. Thus $m_{1}$ is about $2m_{G}$,
and is an unlikely candidate for a gluon mass.

On the other hand, $m_{2}$ is slightly larger than $m_{G}/2$. 
If we view the lightest scalar glueball as being composed
of two consituent gluons, then it is natural to identify $m_2$
as the gluon mass.
Furthermore, $m_{2}$ is the lightest mass in the vector channel, and it is
logical to identify this with the gluon. In the confined phase
interpretation of the unquenched model, the lightest vector is a bound state
of a gluon, scalar, and anti-scalar. In the Higgs phase, the lightest vector
is simply the massive vector particle formed by the Higgs mechanism. We
expect that any meaningful gluon mass would be independent of $\alpha $, and 
$m_{2}$ appears to be approximately constant over the range of $\alpha $
studied.

\begin{figure} 
  \includegraphics[width=5in]{masses_16.eps}
  \caption{Values of the two mass parameters in 
	    as a
	    function of $\alpha$ on the $12^3\times 16$ lattice.
	    The light mass, $m_2$ 
	    is approximately constant as a function of $\alpha$, 
	    while the heavy mass initially decreases with
	    increasing $\alpha$, reaching a somewhat constant value.}
	  \label{fig:masses16}
\end{figure}

\begin{figure} 
  \includegraphics[width=5in]{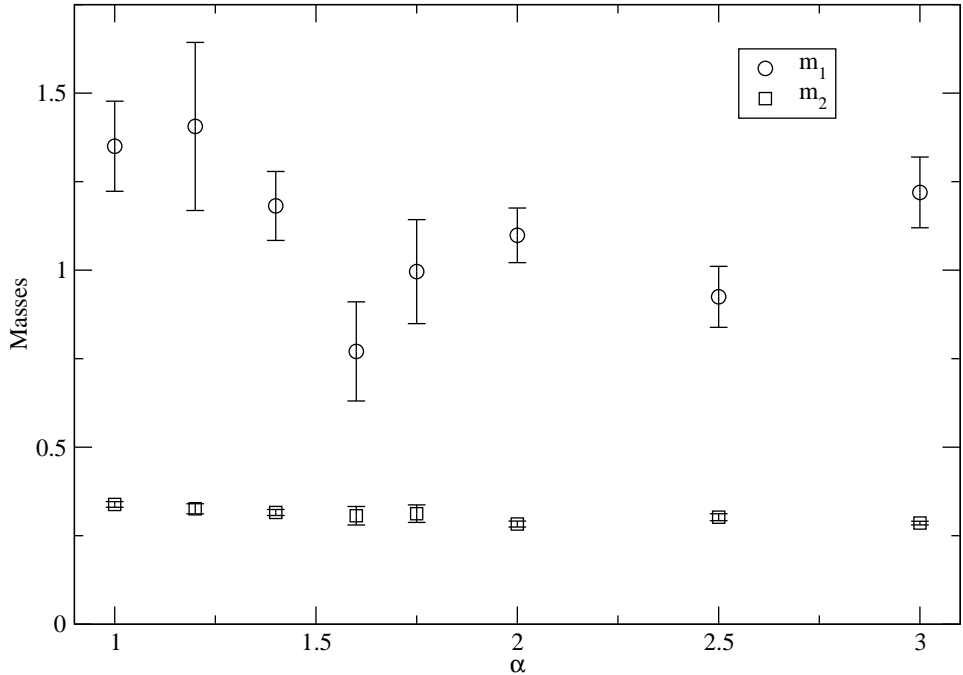}
  \caption{Same as Fig.~\ref{fig:masses16}, but for the $20^4$ lattice.}
	  \label{fig:masses20}
\end{figure}

\section{Discussion}\label{sec:disc}

A large number of possible forms for the gluon propagator, in various
gauges, have been proposed. A brief review is available in
Ref.~\cite{Mandula:nj}. We confine ourselves here to a few remarks on
some of the main themes that have been considered in the literature.
We begin by noting that there is no evidence from lattice simulations
for massless excitations coupled to the gluon propagator. In
particular, a $1/p^{4}$ propagator, which would lead to a confining
potential from one-gluon exchange \cite{Mandelstam:1979xd}, is firmly
ruled out.

The possibility that the gluon propagator might vanish at zero momentum was
first discussed by Gribov, based on his work on gauge 
copies \cite{Gribov:1977wm}. He
proposed that the gluon propagator has complex poles, leading to a gluon
propagator of the form 
\begin{equation}
\frac{p^{2}}{p^{4}+m^{4}}. 
\end{equation}
This leads to the vanishing of the gluon propagator at $p=0$, and a peak in
the gluon propagator at a non-zero momentum. Others have also explored this
possibility and possible generalizations 
\cite{Cucchieri:2000kw,vonSmekal:1997is}.
Zwanziger has proven \cite{Zwanziger:gz} that the gluon propagator
in lattice Landau gauge ($\alpha =\infty $) is zero at $p^{2}=0$ in the
infinite volume limit. His proof does not require the continuum limit, 
but does require that only
configurations in a restricted region of configuration space,
such as the fundamental modular region, be integrated over.
There is evidence from some lattice simulations for a peak in the gluon
propagator at very low wave number 
\cite{Cucchieri:2000kw,Cucchieri:2003di}. However, extremely large lattices
are required, and the propagator does not appear to be trending to zero at
$p^2=0$ as the volume grows.

In our lattice simulations, the gluon propagator does not vanish at $p^{2}=0$
and our best fit to the data does not indicate a peak to non-zero momentum.
However, within the framework of our mixing model, it is possible for a peak
to occur, depending on the values of various parameters. If the gluon
propagator is initially rising with $p^{2}$, it must eventually fall. Thus
the condition for a peak to occur is 
\begin{equation}
2m_{1}^{2}\kappa \lambda \mu ^{2}m_{2}^{2}-2m_{1}^{2}\lambda ^{2}\mu
^{4}-m_{1}^{4}\lambda ^{2}m_{2}^{2}+2m_{2}^{4}\kappa \lambda \mu
^{2}-m_{2}^{2}\lambda ^{2}\mu ^{4}-m_{2}^{6}\kappa ^{2}>0. 
\end{equation}
Our measured parameters are such that this inequality never holds.
We have also fit our data with the form used
by Cucchieri et al.\ in Ref.~\cite{Cucchieri:2000kw}, a generalization 
of the Gribov form. Although a
reasonable fit to the data is obtained, the $\chi ^{2}$ is 50-100 times larger
than that obtained with our quenched mixing form, and a peak is not always
seen. When a peak does appear when fitting to the generalized Gribov form,
it occurs at very small non-zero momenta, directly accessible only on very large
lattices.
We conclude that a peak is a possible feature of the gluon
propagator, but the existence of a peak is sensitive to
the precise form used in fitting.
Such a peak may lack fundamental significance, particularly since it
may only be directly observable on lattices much larger than
the scale on which confinement appears.

Our results for the gluon propagator are similar to those of Leinweber 
\textit{et al.} \cite{Leinweber:1998uu}, who fit $SU(3)$ lattice Landau gauge
propagators to a variety of possible forms. They obtain a best fit with
their model A, for which the propagator has the form 
\begin{equation}
D\left( p^{2}\right) =Z\left[ \frac{AM^{2\alpha_L }}{\left( p^{2}+M^{2}\right)
^{1+\alpha_L }}+\frac{L(p^{2},M^{2})}{p^{2}+M^{2}}\right] 
\end{equation}
where $L$ is a logarithmic factor mimicking one-loop corrections.
Their preferred value for their parameter $\alpha_L $ is
$2.17_{-19}^{+11}$.  This form is thus rather similar to our quenched
mixing form, since for $\alpha_L=2$, it has a triple pole. Of course,
we have two mass scales, $m_{1}$ and $m_{2}$, playing roles in the
denominators.

We have not explored the interesting possibility that the gluon
propagator has an anomalous dimension, first suggested by Marenzoni 
\textit{et al.} \cite{Marenzoni:td}. This behavior could occur 
in our formulation. For example, the
bubble diagram discussed in Sec.~\ref{sec:phase_diag} could give this behavior.
However, high precision data as well as a detailed functional form for
fitting would be needed to test for this behavior.

\section{Conclusions}\label{sec:conc}

We have shown in detail how the quenched character of lattice gauge
fixing leads to violations of spectral positivity in non-Abelian gauge
theories. Although we have focused on lattice Landau gauge, this
problem is likely to occur in any lattice gauge fixing scheme in which
the gauge fixing variables are determined subsequent to the generation
of lattice gauge field configurations. A key step in our analysis has
been the generalization of conventional lattice gauge fixing, where
one configuration is chosen on the gauge orbit, to a formalism
including a gauge-fixing parameter, similar to continuum gauge
fixing. This parameter controls the weighting of configurations along
the gauge orbit, and conventional lattice gauge fixing is formally
recovered as a limiting case. We have not examined lattice gauge
fixing schemes in which the gauge field integration is restricted to a
subset of the entire configuration space, \textit{e.g.}, the
fundamental modular domain, but the relation of these to standard
computational practice is unclear. There are some lattice gauge fixing
schemes, such as lattice Laplacian gauge \cite{Vink:1992ys} which are
defined by a complicated algorithmic process, and may not generalize
in a manner similar to lattice Landau gauge.  Nevertheless, we believe
that the quenched character of these algorithms is the root of
spectral positivity violation there as well.

We have focused here on the form of the gluon propagator, the extraction of
masses, their interpretation and dependence on the gauge-fixing parameter $
\alpha $. The issue of the continuum limit for gluon
propagator masses has not yet been resolved.
In the gauge-invariant sector, physical masses scale as the
gauge coupling $\beta $ is taken to infinity in the manner prescribed by
the renormalization group. In principle, it is also necessary to adjust $
\alpha $ to obtain a continuum limit of gauge-variant quantities.
In other words, the running of the gauge-fixing parameter must be taken into
account. This is similar to the running of the gauge-fixing parameter in covariant
gauges in the continuum. If a meaningful gluon mass, independent of $\alpha $
for large $\alpha $, can be extracted, it might scale correctly in the
continuum limit without tuning $\alpha $. As we have observed, the mass $
m_{2}$ is approximately independent of $\alpha $ for large $\alpha $ at $
\beta =2.6$, and has a value phenomenologically consistent with a
constituent gluon mass. Further study is required to see if $m_{2}$
and $m_{1}$ survive in the continuum limit. Because the standard gluon
operator creates a mixture of states, it is likely that a variational
analysis \cite{Albanese:ds,Teper:wt} using a several operators 
would give a better determination
of $m_{2}$ and particularly $m_{1}$.

We have used the group $SU(2)$ to study gluon properties at zero
temperature. Perhaps the most important application of lattice gauge
fixing is the measurement of electric and magnetic masses in the
deconfined phase.  We are extending our work to finite temperature
$SU(2)$, generalizing the work of
Refs.~\cite{Heller:1995qc,Heller:1997nq,Cucchieri:2000cy} to the case
of finite $\alpha$.  However, it is $SU(3)$ which is of principal
interest \cite{Nakamura:2003pu}. Because gauge fixing can be carried
out any time after lattice field configurations are generated, it is
relatively easy to study gauge fixing on available unquenched $SU(3)$
configurations. In that case, it would be possible to also examine the
dependence of the quark mass on the gauge parameter $\alpha$.

\end{document}